\documentclass[12pt]{article}

\usepackage{graphics,amssymb,epsfig,float}
\usepackage[usenames,dvips]{color}
\usepackage{graphicx}
\usepackage{epsfig}
\usepackage{rotating}
\usepackage{dcolumn}
\usepackage{bm}
\usepackage{cite}
\usepackage{amsmath}

\def\lsim{\;\raise0.3ex\hbox{$<$\kern-0.75em\raise-1.1ex\hbox{$\sim$}}\;}
\def\gsim{\;\raise0.3ex\hbox{$>$\kern-0.75em\raise-1.1ex\hbox{$\sim$}}\;}
\def\beq{\begin{equation}}   \def\eeq{\end{equation}}
\def\ba{\begin{array}}       \def\ea{\end{array}}
\def\bea{\begin{eqnarray}}   \def\eea{\end{eqnarray}}
\def\nn{\nonumber}

\newcommand{\di}{\mathrm{d}}

\begin{document}

\begin{titlepage}
\begin{flushright}
LPT Orsay 12-43
\end{flushright}

\begin{center}
\vspace{1cm}
{\Large\bf Upper Bounds on Asymmetric Dark Matter Self Annihilation
Cross Sections} 
\vspace{2cm}

{\bf{Ulrich Ellwanger and Pantelis Mitropoulos}}
\vspace{1cm}\\
\it  LPT, UMR 8627, CNRS, Universit\'e de Paris--Sud, 91405 Orsay,
France \\
\end{center}

\vspace{1cm}
\begin{abstract}
Most models for asymmetric dark matter allow for dark matter self
annihilation processes, which can wash out the asymmetry at temperatures
near and below the dark matter mass. We study the coupled set of
Boltzmann equations for the symmetric and antisymmetric dark matter
number densities, and derive conditions applicable to a large class of
models for the absence of a significant wash-out of an asymmetry. These
constraints are applied to various existing scenarios. In the case of
left- or right-handed sneutrinos, very large electroweak gaugino masses,
or very small mixing angles are required.
\end{abstract}

\end{titlepage}

\section{Introduction}
In the cosmological Standard Model, the present day baryon density
$\Omega_B$ and the dark matter (DM) relic density $\Omega_{DM}$ are of
similar order of magnitude: $\Omega_{DM}\approx 4.7\times \Omega_B$.
Typically one considers DM in the form of WIMPs (Weakly Interacting
Massive Particles), which are in thermal equilibrium in the very early
universe. Their stability can be ensured by a conserved quantum number
of a discrete $\mathbb{Z}_N$ symmetry, such as $R$-parity in
supersymmetric theories. Once the temperature falls below the WIMP mass,
WIMP self annihilation processes and the expansion of the universe
reduce the WIMP density until the self annihilation rate falls below the
expansion rate of the universe, from where on the comoving WIMP density
remains constant. Assuming that the present day WIMP relic density
coincides with $\Omega_{DM}$ and that its mass is in the typical range
of the order $10-100$~GeV for cold DM and for new particles in models
beyond the Standard Model (BSM), leads to a necessary WIMP self
annihilation cross section of the order of weak interaction cross
sections.

However, this origin of $\Omega_{DM}$ would be completely disconnected
from $\Omega_B$, which is conventionally attributed to a baryon
asymmetry $\eta_B$ originating from CP violating process in the early
universe. Concrete calculations of $\Omega_{DM}$ show that it is quite
sensitive to the WIMP self annihilation cross section and the WIMP mass,
and varies over many orders of magnitude as function of the unknown
parameters in BSM models attempting to explain the present DM relic
density. Then the similar sizes of $\Omega_{DM}$ and $\Omega_B$ result
from a numerical coincidence.

Asymmetric DM (ADM) (see \cite{Nussinov:1985xr,Barr:1990ca,Barr:1991qn,
Dodelson:1991iv,Kaplan:1991ah} for some early discussions, and
\cite{Davoudiasl:2012uw} for a review) is an attempt
to explain the proximity of $\Omega_{DM}$ and $\Omega_B$. The particles
$X$ forming the DM are assumed to be distinct from their antiparticles
$\bar{X}$, and to carry a certain quantum number. The corresponding
charge density of the universe is assumed to be related to baryon number
through equilibrium processes in the early hot universe, such that the
asymmetries $\eta_B$ and $\eta_X$ become related. If the $X -
\bar{X}$ annihilation rate $\sigma_{X\bar{X}}$ is sufficiently
large, the resulting $X$ relic density will be determined exclusively by
the asymmetry $\eta_X$. Then one obtains $\Omega_{DM} \simeq \frac{M_X
\eta_X}{M_p \eta_B} \Omega_B$ (where $M_p$ is the proton mass, and $M_X$
the mass of the $X$ particles), which gives the correct order if $\eta_X
\approx \eta_B$.

In supersymmetric scenarios, the discrete $\mathbb{Z}_N$ symmetry
responsible for the stability of the particles $X$ is equal to or
related to $R$-parity. For instance, sneutrinos (left-handed or
right-handed, mixtures thereof or mixtures with singlets) have been
proposed as ADM \cite{Hooper:2004dc,McDonald:2006if,
Abel:2006nv,Page:2007sh, Kang:2011wb,Ma:2011zm}. Singlet extensions of
the MSSM have been considered in \cite{Suematsu:2005zc,Kaplan:2009ag}.
Recently, higgsinos have been suggested in \cite{Blum:2012nf}, since
they possess a conserved quantum number before the electroweak phase
transition. In all these cases, the dark matter asymmetry can be related
to the baryon and/or lepton asymmetry through sphaleron, gauge and
Yukawa induced process (see \cite{Chung:2008gv} and references therein).
As long as the baryon/lepton asymmetry is generated at temperatures
above the freeze out of the processes which transfer it to the ADM,
this mechanism is independent from the precise CP and baryon/lepton
number violating origin of the baryon asymmetry.

Of course, in the case of sizeable ADM couplings to Higgs or $Z$ bosons,
the direct detection rate must be studied and must not exceed present
bounds \cite{Aprile:2011hi}. (See \cite{MarchRussell:2012hi} for a
discussion of the potential conflict between a sufficiently large $X -
\bar{X}$ annihilation rate and a too large direct detection rate.) A too
large direct detection rate
can be avoided through mass splittings leading to inelastic scattering
\cite{Ma:2011zm,Blum:2012nf}. Alternatively, one may consider that the
$X$ particles -- after their density has frozen out to a value related
to the baryon density -- decay later into other essentially inert
particles with very small couplings, which have not been in thermal
equilibrium. Subsequently we will leave aside the problem of direct
detection rates.

However, typically the discrete $\mathbb{Z}_N$ symmetry responsible for
the stability of the particles $X$ does not forbid $X-X$ self
annihilation processes, through the same couplings which transfer the
baryon or lepton asymmetry to the $X$ particles. Once $X-X$ self
annihilation processes are allowed, these processes can wash out the
asymmetry $\eta_X$.

This conclusion is too naive: A rough condition for the absence of a
wash-out is to require that, at temperatures of the order of the DM
mass, the rate of these processes is below the Hubble expansion rate.
Subsequently we study this phenomenon quantitatively in the form of the
coupled set of Boltzmann equations for the symmetric and antisymmetric
dark matter number densities. We find that the upper bounds on
$\sigma_{XX}$ are extremely strong if one wishes to obtain a final $X$
number density which is dominated by its asymmetry $\eta_X$ such that
$\Omega_{DM}$ is related to $\Omega_B$ as described above.

Interestingly it turns out that, under typical assumptions as a large
$X - \bar{X}$ annihilation rate, a fairly model independent upper
bound on $\sigma_{XX}$ (depending in a simple way on $M_X$ and the
$s$-wave or $p$-wave nature of the $X - X$ annihilation
process) can be derived from the only condition that the $X$ asymmetry
is $\emph not$ reduced by a large amount. The determination of this
upper bound is the main result of this paper. Subsequently we apply it
to sneutrino and higgsino ADM scenarios, and to the singlet extension of
the MSSM proposed in \cite{Kaplan:2009ag} (in an approach similar to,
but slightly different from \cite{Graesser:2011wi}).

Our approach is based on the Boltzmann equations for the $X$ and
$\bar{X}$ number densities in the presence of an $X - \bar{X}$
asymmetry. Boltzmann equations in the presence of asymmetries have been
considered previously e.g. in \cite{Griest:1986yu,Belyaev:2010kp,
Graesser:2011wi,Iminniyaz:2011yp} where, however, the $X-X$ annihilation
rate $\sigma_{XX}$ was assumed to vanish. We find that even a small
$X-X$ annihilation rate $\sigma_{XX}$ can have a strong (negative)
impact on the resulting $X - \bar{X}$ asymmetry.

In the next Section we establish the Boltzmann equation for the $X -
\bar{X}$ asymmetry for non-vanishing $\sigma_{XX}$, and clarify the
assumptions allowing for its model-independent integration. Our main
results are upper bounds on $\sigma_{XX}$ depending on the tolerated
dilution of the initial $X - \bar{X}$ asymmetry. In Section~3 we
study the consequences of this result for sneutrino and higgsino ADM
scenarios. In Section~4 we consider the model with a term
$X^2LH/\Lambda$ in the superpotential introduced in \cite{Kaplan:2009ag}
where the expression for $\sigma_{XX}$ is different from the previous
scenarios. A summary and conclusions are given in Section~5.

\section{Boltzmann equations for asymmetric dark matter}

The common mass of the DM particles $X$ and its antiparticle
$\bar{X}$ is denoted by $m$. Their equilibrium number densities are
assumed to differ by a chemical potential $\mu$. Assuming
Maxwell-Boltzmann statistics (justified for temperatures $T \lsim m/3$),
the equilibrium number densities are given by
\beq\label{eq:1}
n_{X}^{eq}=\frac{T}{2\pi^2}g m^2 \mathrm{K}_2(m/T)\mathrm{e}^{\mu/T}\; ,
\qquad
n_{\bar{X}}^{eq}=
\frac{T}{2\pi^2}g m^2 \mathrm{K}_2(m/T)\mathrm{e}^{-\mu/T}\; .
\eeq
$g$ denotes the number of internal degrees of freedom of $X$, and
$\mathrm{K}_2 $ the modified Bessel function of the second kind. We
assume that the $X - \bar{X}$ asymmetry has been generated during
periods before the one considered here, through processes (such as
sphaleron processes) which have frozen out. Since we assume that the $X
- \bar{X}$ asymmetry is related to the baryon asymmetry, $\mu/T$ is
very small: $\mu/T \lsim 10^{-9}$.

It is convenient to introduce number densities per comoving volume, $Y =
n/s$, with $s$ the entropy density:
\beq\label{eq:2}
 s=h_{eff}(T)\frac{2\pi^2}{45}T^3, 
\end{equation}
where
\begin{equation}\label{eq:3}
 h_{eff}(T)=\sum_{bosons}g_i \left( \frac{T_i}{T} \right)^3
 +\frac{7}{8} \sum_{fermions}g_i \left( \frac{T_i}{T} \right)^3 
\end{equation}
is a sum over the effective massless degrees of freedom. Furthermore one
translates the time dependence of the number densities into a
temperature dependence, using
\beq\label{eq:4}
 H=\left( \frac{8\pi}{3}G \rho \right)^{1/2}
\eeq
for the Hubble parameter in the radiation dominated epoch and
\beq\label{eq:5}
 \rho=g_{eff}(T)\frac{\pi^2}{30}T^4 
\eeq
for the (relativistic) matter density. $g_{eff}$ is defined as in
\eqref{eq:3}
with cubic powers replaced by quartic powers. Assuming a two-body final
state and using the principle of detailed balance \cite{Gondolo:1990dk},
the Boltzmann equations for $Y_{X}$ and $Y_{\bar{X}}$ as functions
of $x\equiv m/T$ become
\beq\label{eq:6}
  \frac{\di Y_X}{\di x}=-\sqrt{\frac{\pi}{45 G}}
\frac{g_*^{1/2}m}{x^2} 
 \left[ 
   \langle \sigma_{XX} v \rangle \left(Y_X^2-Y_{X}^{eq\ 2}\right)
  +\langle \sigma_{X\bar{X}} v \rangle 
  \left(Y_X Y_{\bar{X}}-Y_{X}^{eq} Y_{\bar{X}}^{eq}\right)
 \right]\; ,
 \end{equation}
 \begin{equation}\label{eq:7}
  \frac{\di Y_{\bar{X}}}{\di x}=-\sqrt{\frac{\pi}{45 G}}
\frac{g_*^{1/2}m}{x^2} 
 \left[ 
   \langle \sigma_{XX} v \rangle 
   \left(Y_{\bar{X}}^2-Y_{\bar{X}}^{eq\ 2}\right)
  +\langle \sigma_{X\bar{X}} v \rangle 
  \left(Y_X Y_{\bar{X}}-Y_{X}^{eq}Y_{\bar{X}}^{eq}\right)
  \right]
 \end{equation}
where the effective number of degrees of freedom $g_*$ is given by
\begin{equation}\label{eq:8}
 g_*^{1/2}\equiv\frac{h_{eff}}{g_{eff}^{1/2}}
\left(1+\frac{1}{3}\frac{T}{h_{eff}}\frac{\mathrm{d}h_{eff}}
{\mathrm{d}T} \right)\; .
\end{equation}
Here we have assumed that the self annihilation cross sections satisfy
$\sigma_{\bar{X}\bar{X}}=\sigma_{XX}$, and  $ \left\langle 
\sigma v \right\rangle $ is the thermal average of the cross
section times velocity given by
\begin{equation}\label{eq:9}
 \left\langle  \sigma v \right\rangle = 
 \frac{1}{8m^4 T K_2^2 (m/T)} \int_{4m^2}^\infty{\sigma(s) 
 \left(s-4m^2\right)\sqrt{s}\;
 \mathrm{K}_1 ( \sqrt{s} /T ) \mathrm{d}s}\; .
\end{equation}
In most cases $\left\langle  \sigma v \right\rangle$ can be
expanded in powers of the relative velocity of the incoming particles.
Then, the thermal average is approximated by an expansion in powers of
$1/{x}$:
\begin{equation}\label{eq:10}
 \langle \sigma v \rangle \simeq a + bx^{-1} +
\mathcal{O}\left(x^{-2}\right)\; .
\end{equation}

It is useful to consider the difference and the sum of $Y_X$ and
$Y_{\bar{X}}$ defined by $A$ and $Z$, respectively:
\begin{equation}\label{eq:11}
A=Y_X - Y_{\bar{X}}, \qquad Z=Y_X + Y_{\bar{X}}\; .
\eeq
Then the Boltzmann equation for $A$ becomes
\begin{equation}\label{eq:12}
  \frac{\di A}{\di x}=-\sqrt{\frac{\pi}{45 G}} 
 \frac{g_*^{1/2}m}{x^2} 
 \langle \sigma_{XX} v \rangle \left( Z A-Z^{eq} A^{eq}\right)
 \eeq
where, neglecting terms of ${\cal O}(\mu/T)^2$,
\begin{equation}\label{eq:13}
 Z^{eq}= \frac{45g}{2 \pi^4}\frac{x^2 K_2(x)}{h_{eff}(m/x)}\; , \qquad
 A^{eq}=\frac{\mu}{T} Z^{eq}\; .
 \eeq
For $\sigma_{XX} \ll \sigma_{X\bar{X}}$ we can neglect $\sigma_{XX}$
in the Boltzmann equation for $Z$, which becomes
\begin{equation}\label{eq:14}
\frac{\di Z}{\di x}=-\sqrt{\frac{\pi}{45 G}} \frac{g_*^{1/2}m}{x^2} 
 \langle \sigma_{X\bar{X}} v \rangle 
 \frac{1}{2}\left(Z^2-A^2-Z^{eq\ 2}+ A^{eq\ 2}\right)\; .
 \eeq
Previously these Boltzmann equations were investigated in
\cite{Griest:1986yu,Belyaev:2010kp, Graesser:2011wi,Iminniyaz:2011yp}
under the assumption that $A$ remains constant, i.e. that the right hand
side of (\ref{eq:12}) vanishes. If $\sigma_{X\bar{X}}$ is large
enough, the freeze-out temperature is low, and $Z \sim Z^{eq}$ to a very
good approximation over a long period, and finally $Z_{t\to \infty}\sim
A$ up to corrections studied in \cite{Graesser:2011wi,Iminniyaz:2011yp}.
This is the desired result leading to a DM relic density determined by
$A$ which, in turn, is supposed to be related to the baryon asymmetry.

During the period where $Z \sim Z^{eq}$, (\ref{eq:12}) simplifies to
\begin{equation}\label{eq:15}
  \frac{\di A}{\di x}=-\sqrt{\frac{\pi}{45 G}} 
 \frac{g_*^{1/2}m}{x^2} 
 \langle \sigma_{XX} v \rangle Z^{eq}\left(A- A^{eq}\right)\; ,
 \eeq
which can be integrated with the usual initial condition $A_{in} \sim
A^{eq}$ for $T\sim m$ or $x \sim 1$, and a given expression for $m
\langle \sigma_{XX} v \rangle$. Note that it is $A_{in}$ which is
assumed to be related to the baryon asymmetry. As in the usual case one
finds that $A$ freezes out at a freeze-out temperature $T_f=m/x_f^{A}$,
and $A_\infty \equiv A_{t \to \infty} \sim A^{eq}(x_f^{A})$. However,
since eq.~(\ref{eq:15}) is linear in $A$, the ratio $R\equiv
A_\infty/A_{in}$ is independent from $A_{in}$ and hence independent from
$\mu/T$. The ADM paradigm requires that $R$ is not too small; otherwise
$A_\infty$ is sensitive to $\langle \sigma_{XX} v \rangle$ as in usual
DM scenarios, and $\Omega_{DM}\approx \Omega_B$ remains a numerical
coincidence.

The dominant dependence on the parameters of the model originates from
the combination $m \langle \sigma_{XX} v \rangle$ in
(\ref{eq:15}), an additional weak dependence on $m$ arises from the
effective number of degrees of freedom in $h_{eff}(m/x)$ in
(\ref{eq:13}) and in $g_*$. For these we use the parametrization given
in \cite{Hindmarsh:2005ix}. Now, $R$ can be computed for any given
expression for $m \langle \sigma_{XX} v \rangle$. (We solved the
coupled set (\ref{eq:6},\ref{eq:7}) of Boltzmann equations, but for
$\sigma_{XX} \lsim 10^{-5}\sigma_{X\bar{X}}$ we have $Z \simeq
Z^{eq}$ for the relevant range of $x$, and the result for $R$ is
independent from $\sigma_{X\bar{X}}$ as it is obvious from
(\ref{eq:15}).) Assuming eq.~(\ref{eq:10}), we will consider the two
cases (a) $\langle \sigma_{XX} v \rangle \simeq a$ ($s$-wave
annihilation) and (b) $\langle \sigma_{XX} v \rangle \simeq b/x$
($p$-wave annihilation; in practice one may find a combination of both).
The results for $R$ as function of $\log\left(m a \right)$ and
$\log\left(m b \right)$ are shown in Fig.~\ref{fig:1}.

\begin{figure}
\begin{center}
 \includegraphics[width=0.7\textwidth]{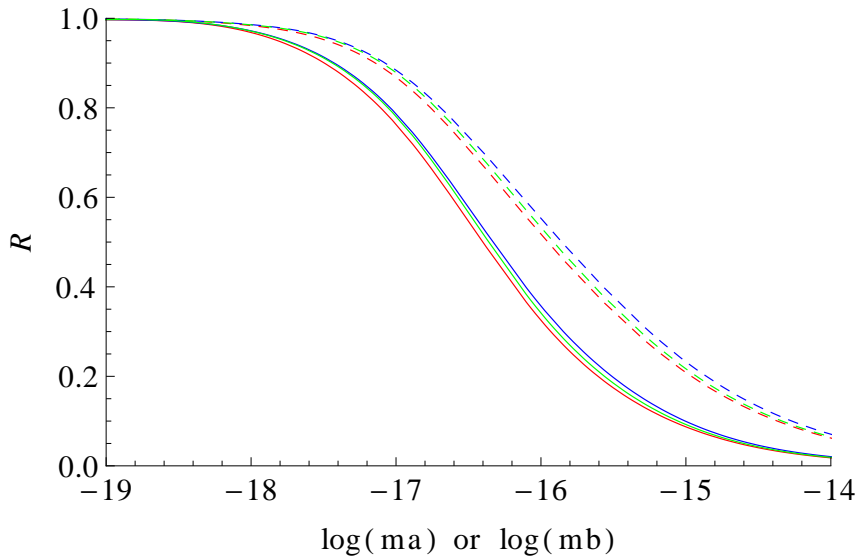}
\end{center}
 \caption{The ratio $R\equiv A_\infty/A_{in}$ of the final asymmetry
over the initial asymmetry as a function of $\log\left(ma \right)$ or
$\log\left(ma \right)$, with $ma$ or $mb$ in GeV. The solid lines
represents the values of $R$ if $\langle \sigma_{XX} v \rangle = a$,
and the dashed lines the values of $R$ if $\langle \sigma_{XX} v
\rangle = b/x$. The color corresponds to the mass of the DM particle,
red for $m=10 \, GeV$, green for $m=100 \, GeV$ and blue for $m=1 \,
TeV$.}
 \label{fig:1}
\end{figure}

We see that within the cases (a) or (b) the dependence on $m$ beyond the
one in $m \langle \sigma_{XX} v \rangle$ is negligibly small, and
we can deduce upper bounds on $m \langle \sigma_{XX}
v \rangle$ as function of the tolerated reduction $R$ of the
asymmetry:
\bea\label{eq:16}
R > \text{0.5:}& m a \lsim 5\times10^{-17}\;\text{GeV}^{-1}\ \text{(case
(a))}, \ m b \lsim 1\times10^{-16}\;\text{GeV}^{-1}\ \text{(case (b))}\;
,\nn \\
R > \text{0.1:}& m a \lsim 1\times10^{-15}\;\text{GeV}^{-1}\ \text{(case
(a))},\ m b \lsim 5\times10^{-15}\;\text{GeV}^{-1}\ \text{(case (b))}\; .
\eea

Clearly, if we require only a moderate reduction of the asymmetry $A$,
the freeze-out temperature $T_f^A=m/x_f^{A}$ must not be far below $m$,
or $x_f^{A}$ must not be too large. (Here we define $x_f^{A}$ as the
temperature where the expansion rate of the universe becomes larger than
the rate of self-annihilation, i.e. $H(x_f^{A})=
(n_{X}^{eq}-n_{\bar{X}}^{eq}) \langle \sigma_{XX} v \rangle(x_f^{A})$
which is solved numerically.) In Fig.~\ref{fig:2} we show $x_f^{A}$ as
function of $\log\left(m a \right)$ and $\log\left(m b \right)$ and,
indeed, $x_f^{A}$ is well below 5 if the first of the conditions
\eqref{eq:16} is satisfied.

The requirement of a not too large value of $x_f^{A}$ and the fact that
$\langle \sigma_{XX} v \rangle(x_f^{A}) \sim \mathrm{e}^{x_f^{A}}$
explains why we obtain $\langle \sigma_{XX} v \rangle \ll\langle
\sigma_{X\bar{X}} v \rangle$; the desired value of $x_f$ in case of
$X-\bar{X}$ annihilation is rather of ${\cal O}(20)$.

\begin{figure}
\begin{center}
 \includegraphics[width=0.7\textwidth]{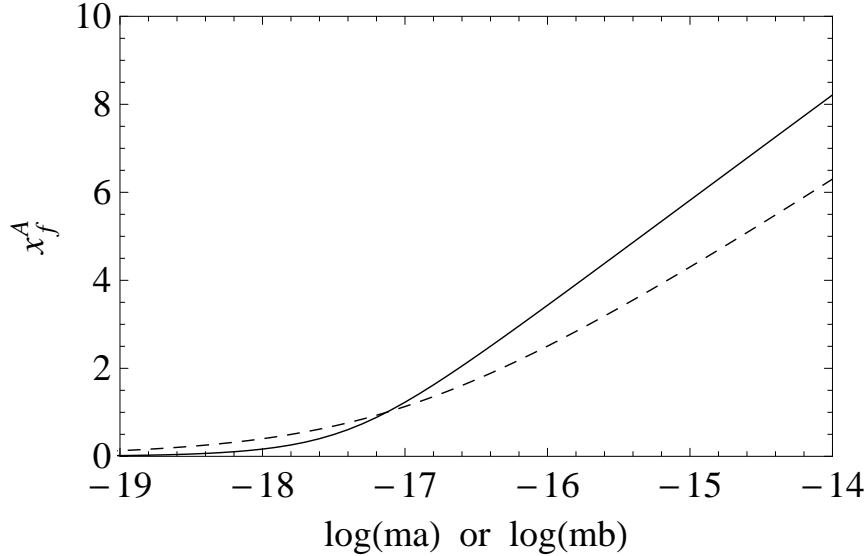}
\end{center}
 \caption{The freeze-out point $x_f^{A}=\frac{m}{T_f^A}$ as a function of
 $\log\left(m a \right)$ (solid line) and $\log\left(m b \right)$
 (dotted line).}
 \label{fig:2}
\end{figure}

In the next Sections we study the implications of the upper bounds on
$\sigma_{XX}$ for various models for supersymmetric ADM.

\section{Sneutrino and higgsino ADM}

Left-handed sneutrinos or mixtures of left- and right-handed sneutrinos
(or singlets) have been proposed as ADM in \cite{Hooper:2004dc,
McDonald:2006if, Abel:2006nv,Page:2007sh, Kang:2011wb,Ma:2011zm}.
Clearly, left-handed sneutrinos $\tilde{\nu}_L$ can self-annihilate
through processes of the form $\tilde{\nu}_L +\tilde{\nu}_L \to{\nu}_L
+{\nu}_L$ by the exchange of electroweak gauginos in the
$t$-channel. Electroweak gauginos are the binos with mass $M_1$, and
winos with mass $M_2$. The corresponding expression for  $\langle
\sigma_{\tilde{\nu}_L\tilde{\nu}_L} v \rangle$ can be obtained from
\cite{Nihei:2002sc}, and reads in the limit $M_1,\ M_2 \gg m$
\beq\label{eq:17}
\langle \sigma_{\tilde{\nu}_L\tilde{\nu}_L} v \rangle \simeq 
\frac{g_2^4}{16\pi} \left(1-\frac{3}{2x}\right) 
\left(\frac{\tan^2\theta_w}{M_1}+\frac{1}{M_2}\right)^2
 \end{equation}
where $g_2$ is the SU(2) gauge coupling and $\theta_w$ the weak mixing
angle. If one assumes universal gaugino masses at the GUT scale, $M_1$
and $M_2$ are related by $M_1\simeq M_2/2$ and, with $\tan^2\theta_w
\approx 0.3$, \eqref{eq:17} simplifies to
\beq\label{eq:18}
\langle \sigma_{\tilde{\nu}_L\tilde{\nu}_L} v \rangle \simeq 
\frac{g_2^4}{8\pi} \left(1-\frac{3}{2x}\right) \frac{1}{M_2^2}\; .
 \end{equation}
From Fig.~\ref{fig:1}, the first term leads to stronger constraints, and
applying the conservative bound $R>0.1$ (case (a)) from \eqref{eq:16}
leads to
\beq\label{eq:19}
M_2 \gsim 3\times 10^7\; \text{GeV}\times 
\left(\frac{m}{100\;\text{GeV}}\right)^{1/2}
\eeq
which excludes gaugino masses of the order of the electroweak scale.
(Even for such large gaugino masses the sneutrino-antisneutrino
annihilation rate would remain large due to processes with slepton
exchange in the t-channel.)

If the ADM $X$ consists in a mixture of left-handed sneutrinos
$\tilde{\nu}_L$ and right-handed sneutrinos or other electroweak
singlets, the result depends on the $\tilde{\nu}_L$ component of $X$ or
the mixing angle $\sin\delta$ where $X =\tilde{\nu}_L \sin\delta +
\dotsc$:
\beq\label{eq:20}
\langle \sigma_{XX} v \rangle \simeq 
\frac{g_2^4\sin^4\delta}{8\pi}\left(1-\frac{3}{2x}\right)
\frac{1}{M_2^2}\; .
 \end{equation}
Now the same argument leads to
\beq\label{eq:21}
\sin^2\delta \lsim 3.3\times 10^{-6} \times
 \frac{M_2}{\sqrt{m\cdot 100\;\text{GeV}}}\; .
\eeq
Hence, for $M_2\approx m \approx 100$~GeV, $\sin\delta$ has to be very
small independently from constraints from direct DM detection
experiments -- or, for mixing angles $\sin\delta \approx 1$, one is lead
back to~\eqref{eq:19}.

Higgsinos $\tilde{h}_u$, $\tilde{h}_d$ with masses $m \sim 200-1000$~GeV
as ADM were proposed recently in \cite{Blum:2012nf}. Before the
electroweak phase transition (where the Higgs vevs develop), higgsinos
can be considered as mass eigenstates. Through sphaleron, gauge and
Yukawa interactions at high temperature, a higgsino asymmetry  $
A_{\tilde{h}_u}\sim A_{\tilde{h}_d}$ proportional to the baryon 
asymmetry is generated \cite{Chung:2008gv,Blum:2012nf}. If the
higgsinos  are the LSPs (lightest supersymmetric particles) and other
sparticles are sufficiently heavy, this asymmetry could survive until
today \cite{Blum:2012nf} provided the $\tilde{h}_u \tilde{h}_u$ (and
$\tilde{h}_d \tilde{h}_d$) self annihilation rates are sufficiently
small.

However, higgsinos have the same couplings to electroweak gauginos as
sneutrinos, and can again self-annihilate into $H_u$, $H_d$ (to be
considered as eigenstates before the electroweak phase transition)
through $t$-channel exchange of binos and winos. This time the
scattering process is a $p$-wave process (case (b) in \eqref{eq:16})
and we obtain from \cite{Nihei:2002ij} (again for $M_1,\ M_2 \gg m$)
\beq\label{eq:22}
\langle \sigma_{\tilde{h}_u\tilde{h}_u} v \rangle =
\langle \sigma_{\tilde{h}_d\tilde{h}_d} v \rangle \simeq 
\frac{3 g_2^4}{x 8\pi}\left(\frac{\tan^2\theta_w}{M_1} +
\frac{1}{M_2}\right)^2\; .
\eeq
Applying again the conservative bound $R>0.1$ (case (b)) from
\eqref{eq:16} leads to
the same constraint on $M_2$ as in \eqref{eq:19},
which is close to the estimated bound on gaugino masses given in
\cite{Blum:2012nf} from the same argument.

\section{The $\Delta W \sim XXHL/\Lambda$ model}

Left-handed sneutrinos and higgsinos as ADM tend to violate bounds on
direct DM detection cross sections (unless mass splittings are
introduced leading to inelastic scattering
\cite{Ma:2011zm,Blum:2012nf}). Moreover, as we have seen in the previous
section, constraints from sufficiently small self annihilation cross
sections are very strong. These constraints would be alleviated, if the
asymmetry is transferred to lighter essentially inert particles. A
simple model of that kind has been proposed in \cite{Kaplan:2009ag},
where a gauge singlet scalar superfield $X$ and a superpotential
\beq\label{eq:24}
\Delta W = \frac{1}{\Lambda}XX H_u L_i
\eeq
are introduced. (Here $H_u$ denotes a Higgs superfield, and $L_i$ any
left-handed lepton superfield.) This nonrenormalizable interaction can
originate from integrating out heavy vector-like sterile neutrinos or
electroweak doublets \cite{Kaplan:2009ag} with mass $\sim \Lambda$,
typically $\gsim 1\,\mathrm{TeV}$. Another singlet
chiral superfield $\overline{X}$ should be introduced to allow for a
supersymmetric Dirac mass term $M_{X}$ for the fermionic components
$\psi_X \psi_{\overline{X}}$, e.g. via a NMSSM-like singlet $S$ with
$\langle S \rangle \neq 0$ and a coupling $\lambda^\prime S X
\overline{X}$ in the superpotential.

$\psi_{\bar{X}}$, $\psi_X$ would carry lepton number $\pm 1$,
respectively. The superpotential \eqref{eq:24} breaks the usual R
parity, but preserves a $Z_4$ symmetry which allows for the decay of the
usual LSP into $\psi_X\psi_X$ \cite{Kaplan:2009ag}. At high
temperature where the processes induced by \eqref{eq:24} are in
equilibrium together with  sphaleron, gauge and Yukawa interactions,
these imply an asymmetry $A_X$ of $\sim 35\%$ of the baryon asymmetry
\cite{Kaplan:2009ag,Graesser:2011wi}, the precise value depending on
whether top quarks and squarks are still in equilibrium when the
interactions from \eqref{eq:24} decouple. Assuming a sufficiently rapid
$\psi_X - \bar{\psi}_X$ annihilation rate and $M_{X}\sim 11-13$~{GeV},
the relic density is then automatically of the correct order.

After electroweak symmetry breaking, the superpotential \eqref{eq:24}
gives rise to an interaction of the form
\beq\label{eq:25}
\frac{v_u}{\Lambda}\psi_X\psi_X \tilde{\nu}_i\; .
\eeq
(Subsequently we omit the neutrino/sneutrino index $i$.) The sneutrino
$\tilde{\nu}$ does not have to be the LSP; the LSP can be
the lightest neutralino $\tilde{\chi}$. Then an on-shell sneutrino
$\tilde{\nu}$ would decay via the usual vertex $g \tilde{\nu}
\tilde{\chi} \nu$ -- where $g$ is of the order of electroweak gauge
couplings, if $\tilde{\chi}$ is dominantly bino-like -- into
$\tilde{\chi}$ plus a neutrino $\nu$. At energies below the sneutrino
mass $m_{\tilde{\nu}}$, integrating out the sneutrino leads to an
effective four Fermi interaction
\beq\label{eq:26}
\frac{gv_u}{m_{\tilde{\nu}}^2\Lambda}\psi_X\psi_X \tilde{\chi} \nu\; .
\eeq

At energies above $M_{\tilde{\chi}}$, \eqref{eq:26} allows for the
scattering process $\psi_X\psi_X \rightarrow \tilde{\chi} {\nu}$.
However, assuming $M_{\tilde{\chi}}>2 M_{X}$, the lightest neutralino
$\tilde{\chi}$ is not stable. Since $\tilde{\chi}$ is a Majorana
Fermion, \eqref{eq:26} leads to its decay into $\psi_X \psi_X \nu$ and
$\bar{\psi}_X \bar{\psi}_X \bar{\nu}$ with corresponding
branching ratios of $50\%$. The latter case leads to the scattering
process
\beq\label{eq:27}
\psi_X\psi_X \rightarrow \bar{\psi}_X \bar{\psi}_X \bar{\nu}
\bar{\nu}\; .
\eeq

As in the case of sneutrinos and higgsinos, this ADM self annihilation
process can have potentially disastrous consequences for the remaining
asymmetry. In \cite{Kaplan:2009ag}, the rate for this process has been
estimated by integrating out both the sneutrino $\tilde{\nu}$ and the
lightest neutralino $\tilde{\chi}$ with the result that, for
$m_{\tilde{\nu}} \sim M_{\tilde{\chi}} \sim 100$~GeV and $\Lambda \gsim
1$~TeV, it would go out of equilibrium (drop below the Hubble expansion
rate) for decoupling temperatures $T_D$ somewhat above $M_{X}$, in which
case the asymmetry would hardly be washed out.

A more detailed analysis of the ADM self annihilation processes has been
performed in \cite{Graesser:2011wi}. There it was pointed out that, for
$m_{\tilde{\nu}} \sim M_{\tilde{\chi}} \sim 100$~GeV and $\Lambda \gsim
1$~TeV, the dominant ADM self annihilation process is \textit{real}
$\tilde{\chi}$ production through the interaction \eqref{eq:26}, and the
corresponding cross section was given.

Subsequently the authors in \cite{Graesser:2011wi} estimated $T_D$ by
equating the ADM self annihilation rate with the Hubble expansion rate.
They tolerated a considerable wash-out of the asymmetry and/or a
Boltzmann suppression for $T_D<M_{X}$, and studied the necessary
relations between the final values for $A_\infty$ and $Z_\infty$ defined
in \eqref{eq:11} (or $ r_\infty=(Z_\infty-A_\infty)/(Z_\infty+A_\infty)
$), the DM mass $M_{X}$ and the lightest neutralino mass
$M_{\tilde{\chi}}$ required for a desired DM relic density. Clearly, in
most of this parameter space (after a considerable wash-out of the
asymmetry and/or Boltzmann suppression), the desired DM relic density is
no longer simply related to the baryon asymmetry in contrast to ADM
paradigm.

Here we ask the question under which conditions this does \textit{not}
happen, i.e. under which conditions the initial asymmetry $A_X$
determines essentially the DM relic density. As before we assume that
the $\psi_X - \bar{\psi}_X$ annihilation rate is sufficiently large,
such that we can assume $Z\sim Z^{eq}$ in \eqref{eq:12} leading to 
\eqref{eq:15}.

If the threshold for the process \eqref{eq:27} would be $s>4
M_{X}^2$, we could apply our previous formulas. However, for the
(dominant) annihilation process via \textit{real} $\tilde{\chi}$ (+
neutrino) production, the threshold is $s>M_{\tilde{\chi}}^2>4
M_{X}^2$. As a consequence the thermal average of the cross section
times velocity in \eqref{eq:15} depends in a more complicated way on
$M_{X}$, $M_{\tilde{\chi}}$ and notably on the temperature or
$x=M_{X}/T$ such that the expansion \eqref{eq:10} is no longer
applicable. First, we use the cross section $\sigma_{XX}(s)$ for the
process $\psi_X\psi_X \rightarrow \tilde{\chi} \bar{\nu}$ from
\cite{Graesser:2011wi}:
\beq\label{eq:28}
\sigma_{XX}(s)=\frac{\kappa^2 M_{\tilde{\chi}}^4}{256\pi} \frac{1}{s}
\left(\frac{s}{M_{\tilde{\chi}}^2}-1\right)^2
\eeq
where $\kappa=\frac{g v_u}{\Lambda m_{\tilde{\nu}}^2}$. Then, using
\eqref{eq:9} with $M_{\tilde{\chi}}^2$ as lower threshold of the
integral (without expanding in $M_{X}/M_{\tilde{\chi}}$ as in
\cite{Graesser:2011wi}), we obtain
\beq\label{eq:29}
\langle \sigma_{XX} v \rangle = \frac{\kappa^2}{16\pi}
\left( \frac{M_{\tilde{\chi}}}{M_{X}} \right)^2
\frac{(M_{\tilde{\chi}}^2-4M_{X}^2) \mathrm{K}_2
(\frac{M_{\tilde{\chi}}}{M_{X}} x)+\frac{6 M_{\tilde{\chi}}
M_{X}}{x} \mathrm{K}_3(\frac{M_{\tilde{\chi}}}{M_{X}} x)}{x^2
\mathrm{K}_2^2(x)}\; .
\eeq
For $ M_{X} \ll M_{\tilde{\chi}} $, $\langle \sigma_{XX} v
\rangle$ is proportional to $M_{\tilde{\chi}}^{7/2}
\mathrm{e}^{-M_{\tilde{\chi}}/T}$ like the collision term evaluated in
\cite{Graesser:2011wi}. The Boltzmann suppression $\sim
\mathrm{e}^{-M_{\tilde{\chi}}/T}$ is an obvious consequence of the
threshold $s>M_{\tilde{\chi}}^2$ required for real $\tilde{\chi}$
production. Subsequently we integrate the Boltzmann equation
\eqref{eq:15} numerically employing \eqref{eq:29} for $\langle
\sigma_{XX} v \rangle$, which allow us to study $R\equiv
A_\infty/A_{in}$ as before. Now, however, $R$ depends in a more
complicated way on the parameters $\kappa$, $M_{\tilde{\chi}}$ and
$M_{X}$ of the model. On the other hand, the initial asymmetry is
quite well known in this class of models, $A_{in}\sim 0.35\,B$ (where
$B$ is the baryon asymmetry), and finally we must obtain
\beq\label{eq:30}
\frac{\Omega_{DM}}{\Omega_B}=\frac{A_\infty M_{X}}{B m_p}\sim 4.7
\eeq
which determines $M_{X}$ in terms of $A_\infty$ or $R$:
\beq\label{eq:31}
M_{X}\sim (12.5\ \mathrm{GeV})/R\; .
\eeq
Hence, once the correct ADM relic density is imposed, the free
parameters of the model are $\kappa=\frac{g v_u}{\Lambda
m_{\tilde{\nu}}^2}$, $M_{\tilde{\chi}}$ and $M_{X}$ or $R$. In order to
clarify the correlations between these parameters, we show $R$ in the
range $R=1 \ldots 0.1$ ($M_{X}=12.5 \ldots 125$~GeV) as function of
$M_{\tilde{\chi}}$ for various values of $\kappa \leq 10^{-5}$~GeV$^{-2}$
in Fig.~\ref{fig:3}. (Note that all curves continue horizontally along
$R=1$ beyond their upper end.)

\begin{figure}
\begin{center}
 \includegraphics[width=0.7\textwidth]{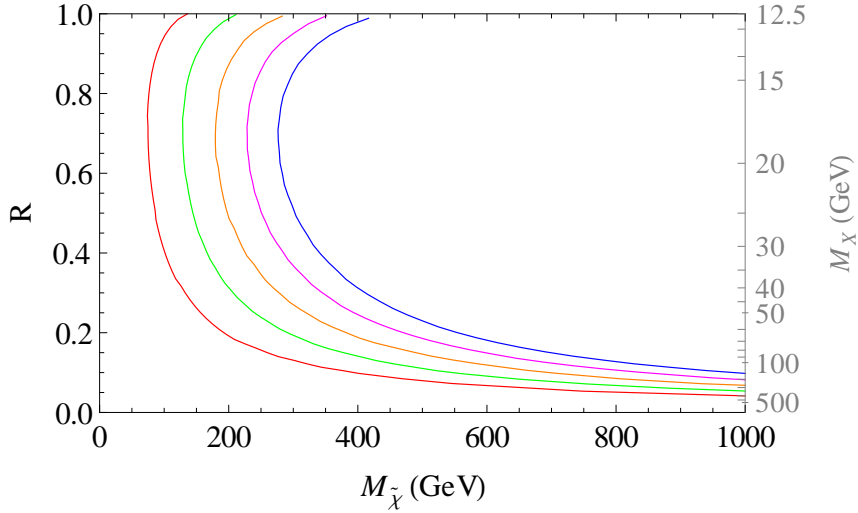}
\end{center}
\caption{$R$ as function of $M_{\tilde{\chi}}$ for different values of
 $\kappa$: red: $\kappa=10^{-9}$~GeV$^{-2}$, green:
 $\kappa=10^{-8}$~GeV$^{-2}$, orange: $\kappa=10^{-7}$~GeV$^{-2}$,
 magenta: $\kappa=10^{-6}$~GeV$^{-2}$, blue:
 $\kappa=10^{-5}$~GeV$^{-2}$. The mass $M_{X}$ of the dark matter,
 related to $R$ as in \eqref{eq:31}, is indicated on the right vertical
 axis.}
\label{fig:3}
\end{figure}

The shape of these curves can be understood as follows: Near the
upper end, $R\sim 1$, $M_{X}\sim 12.5$~GeV and practically no
wash-out takes place by construction, since $\langle \sigma_{XX} v
\rangle$ is too small. For fixed $\kappa$, $\langle \sigma_{XX} v
\rangle$ increases for decreasing $M_{\tilde{\chi}}$ due to  $\langle
\sigma_{XX} v \rangle \sim \mathrm{e}^{-M_{\tilde{\chi}}/T}$, and below
some critical value of $M_{\tilde{\chi}}$, $\langle \sigma_{XX} v
\rangle$ is large enough such that $R$ starts to decrease.

If a sizeable wash-out takes place ($R \lsim 0.5$), it stops when the
annihilation rate falls below the Hubble expansion rate, i.e. for a
given value of $\langle \sigma_{XX} v \rangle$. For fixed $\kappa$, this
implies a certain fixed value for $\mathrm{e}^{-M_{\tilde{\chi}}/T_D}$
or $M_{\tilde{\chi}}/T_D$. Using $x_D=M_{X}/T_D$ and \eqref{eq:31} one
easily derives $R \sim 1/(x_D M_{\tilde{\chi}}$), which explains the
decrease of $R$ for increasing $M_{\tilde{\chi}}$ in this range.

Fig.~\ref{fig:3} allows to identify the regions in parameter space in
which $R$ is not too small ($R \gsim 0.5$), i.e. where $\Omega_{DM}$
follows naturally from $\Omega_B$ as in the ADM paradigm. If the
coefficient $\kappa$ relatively is large, $\kappa \sim
10^{-5}$~GeV$^{-2}$ or $m_{\tilde{\nu}} \sim 100$~GeV and $\Lambda \sim
1$~TeV, one needs $M_{\tilde{\chi}} \gsim 300$~GeV. If $\tilde{\chi}$ is
lighter, $M_{\tilde{\chi}} \sim 100$~GeV, one needs a very small value
of $\kappa \lsim 10^{-9}$~GeV$^{-2}$, hence correspondingly large values
of $m_{\tilde{\nu}}$ and/or $\Lambda$.

\section{Conclusions}

If the dark matter asymmetry is related to the baryon (or lepton)
asymmetry, some couplings must necessarily relate these sectors. The
same couplings can lead to dark matter self annihilation processes,
which can wash out the corresponding asymmetry $A$. A rough condition
for the absence of a wash-out is to require that, at temperatures of the
order of the DM mass, the rate of these processes is below the Hubble
expansion rate. In Section~2 we have studied the corresponding set of
Boltzmann equations quantitatively (assuming a two-body final state).
Requiring a modest wash-out ($A_{\infty}/A_{in} \gsim
0.1$), the upper bounds on  $m \langle \sigma_{XX} v \rangle$ are very
strong, and can be deduced from Fig.~\ref{fig:1} if  $\langle
\sigma_{XX} v \rangle \sim a$ or $\langle \sigma_{XX} v \rangle\sim
b/x$.

If the ADM consists in sparticles with couplings to electroweak gauginos
(left-handed sneutrinos or higgsinos), it follows that the electroweak
gauginos must be extremely heavy such that supersymmetry does not solve
the hierarchy problem. If the ADM consists in particles like
right-handed sneutrinos which mix weakly with left-handed sneutrinos,
the mixing angle must be very small, see eq.~\eqref{eq:21}.

In different models for ADM, the dominant ADM self annihilation process
may be kinematically possible only for $s$ above a threshold larger than
$(2 M_X)^2$, in which case it becomes Boltzmann suppressed and
eq.~\eqref{eq:10} is no longer valid. An example is the popular
$\Delta W \sim XXHL/\Lambda$ model of \cite{Kaplan:2009ag}. Also in this
case, the numerical integration of the Boltzmann equation for the
asymmetry allowed us to specify the range of parameters where the
wash-out of the asymmetry remains modest.

In the past, the constraints following from the absence of a wash-out of
the asymmetry have sometimes been neglected or underestimated (notably
in the case of sneutrinos); we hope that the present work helps to
clarify the relevance of ADM self annihilation processes and the
resulting conditions on corresponding models.

\section*{Acknowledgements}
U.~E. acknowledges partial support from the French ANR LFV-CPV-LHC, ANR
STR-COSMO and the European Union FP7 ITN INVISIBLES (Marie Curie
Actions, PITN- GA-2011- 289442). P.~M. acknowledges support from the
Greek State Scholarship Foundation.


\end{document}